\documentclass[reprint, aps,prd,twocolumn,showpacs,showkeys, 10pt, longbibliography, nolinenumbers, floatfix]{revtex4-2}
\usepackage{amsmath}
\usepackage{graphicx}
\usepackage{float}
\usepackage{dcolumn}
\usepackage{multirow}
\usepackage{natbib}
\usepackage{changes}
\usepackage[colorlinks = true,linkcolor = blue,urlcolor  = blue,citecolor = blue,anchorcolor = blue]{hyperref}
\usepackage{enumerate}
\usepackage{verbatim}
\usepackage{orcidlink}
\usepackage{titlesec}
\usepackage{subcaption}
\usepackage{placeins}

\titlespacing*{\section}{0pt}{\baselineskip}{0pt}
\titlespacing*{\subsection}{0pt}{\baselineskip}{0pt}

\begin{document}

\title{ Classification of Compact Stars via Machine Learning and Neural Network Models}

\author{D. Neraki}

\author{G. Koufetidis}

\author{I. Stergakis}

\author{Th. Diakonidis}

\author{Ch.C. Moustakidis}


\affiliation{Department of Theoretical Physics, Aristotle University of Thessaloniki, 54124 Thessaloniki, Greece  }

\begin{abstract}
In recent years, significant progress has been achieved in the discovery and study of compact astrophysical objects. This advancement has been greatly facilitated by the detection of gravitational waves originating from the mergers of such systems. However, despite these developments, it is still not possible to obtain clear and robust information on the internal structure and composition of these objects.
According to current theoretical considerations, these compact objects may correspond either to neutron stars, where matter composed primarily of neutrons and protons dominates, or to stars consisting of deconfined quark matter, or even to hybrid configurations combining features of both scenarios. At the same time, several theoretical models suggest that these objects may also contain exotic forms of matter, including hyperons, kaon or pion condensates, and possibly even dark matter. In the present study, we attempt to address the question of whether, based on observational measurements of the fundamental properties of compact objects, such as their mass, radius, and tidal deformability, it is possible to reliably infer their internal composition. This investigation is carried out using machine learning based statistical methods. To carry out this study, a large dataset of equations of state has been constructed, including models suitable for describing neutron stars as well as models appropriate for quark stars. Finally, this dataset is employed to generate the corresponding mass–radius  relations, covering the widest possible range of masses and radii. Subsequently, machine learning and deep learning techniques are applied to analyze these configurations. The main conclusion is that an appropriate combination of parameters allows the nature of the compact object to be identified with very high accuracy. Nevertheless, a more systematic investigation is required, taking into account all the possible scenarios mentioned above, before the reliability of the machine-learning–based approach can be firmly established.


\end{abstract}

\maketitle

\section{Introduction}
The study of the structure, composition, and fundamental properties of compact astrophysical objects has become one of the most important areas of contemporary scientific research~\cite{Shapiro:1983du,Haensel2007NeutronS1,Glendenning-2000,schaffner-bielich_2020}. The information that can be extracted through an appropriate combination of astrophysical observations and theoretical modeling is crucial for advancing our understanding of nuclear matter under extreme conditions of density and temperature. Furthermore, given the limitations of terrestrial experiments in probing dense nuclear matter, a significant part of our expectations has been placed on the analysis and study of gravitational waves produced during the merger of compact-object binary systems. The information obtained from such observations complements the already substantial body of knowledge derived from studies of isolated neutron stars as well as neutron stars and white dwarfs participating in binary systems.
The two principal classes of theoretically predicted compact objects are neutron stars and quark stars, while a third related category consists of hybrid stars, which combine structural and compositional characteristics of the former two. In the first category, namely neutron stars, one may consider that, in  addition to baryons, such as neutrons and protons, other degrees of freedom may also become relevant, including hyperons, mesons (such as kaons and pions), and other related particles. In any case, the equation of state (EoS) describing their matter is highly sensitive to its underlying composition. In the case of quark stars, the situation is comparatively simpler, as their composition is assumed to consist exclusively of deconfined quark matter (up, down, and strange quarks). What may vary between different models is the specific form of the interaction among these quarks. Although the composition of these two classes of compact objects is fundamentally different, the similarities in the resulting equations of state that describe them lead to mass–radius configurations with significant overlap, particularly for masses larger than one solar mass. Consequently, a compact object with a given mass and radius may be predicted by both purely baryonic equations of state and equations of state based on deconfined quark matter. As a result, an astrophysical observation, even when the mass and radius of a compact object are measured with relatively good accuracy, may not be sufficient to determine unambiguously to which class the observed object belongs.

Beyond the interplay between theory and observation, a third avenue for advancing our understanding of neutron stars has emerged, namely an integrated approach that combines both. In this context, statistical methods have proven particularly powerful in recent years, with techniques such as Bayesian inference and machine learning (ML) being applied with considerable success~\cite{Read-2009,raithel2016neutron,Ozel-2009,Jiang-2023,Zhou-2023,Huth-2022,Peter-2024,Patra-2025,Imam-2024,Patra-2022,Imam-2022,Clement-2025,Fujimoto-2018,Fujimoto-2020,Thakur-2024,Krastev-2023,Guo-2024,Chatterjee-2024,Zhou-2023b,Soma-2024,Li-2025,Bejger-2025,Papigkiotis-2025,Ventagli-2025,Fujimoto-2024,Carvalho-2024a,Carvalho-2024b,Brandes-2024,Fujimoto-2021,Ferreira-2021,Thakur-2025,Han-2021,Han-2023,Ventagli-2024,Ling-2024,Stergakis-2026}.

Although the body of work devoted to the study of compact objects using various statistical methods is quite extensive, only a limited number of studies have addressed the application of these methods to the classification of compact objects.
Representative studies in this direction include Ref.~\cite{Pattnaik-2021}, in which a ML approach is employed to classify low-mass X-ray binaries according to the nature of their compact object, Ref.~\cite{deBeurs-2022}, which presents a comparative analysis of ML methods for X-ray binary classification and, more closely related to the present work, Ref.~\cite{Carvalho-2024b}, where a ML  framework is developed to detect the presence of hyperons in neutron stars.

In particular, to the best of our knowledge, there are no existing studies that focus on the identification and discrimination between neutron stars and quark stars.
Motivated by the importance of identifying compact astrophysical objects, as well as by the relative scarcity of such studies employing statistical methods, in this work we perform a systematic study, employing machine and deep learning techniques, in order to classify the two main types of compact objects. The classification is based on key physical parameters such as their mass, radius, central pressure, Love number, and tidal deformability. More specifically, we investigate which combination of these parameters is most decisive for the accurate classification of the aforementioned compact objects.

To carry out this study, a large dataset of equations of state has been constructed, together with the corresponding mass–radius configurations derived from them. In particular, as far as neutron stars are concerned, we employ the method described in Refs.~\cite{Read-2009,raithel2016neutron,Stergakis-2026} to construct a comprehensive dataset of equations of state for neutron star cores based on polytropic parameterizations. These equations of state cover a wide region of the pressure–energy density plane, thereby incorporating as much information as possible about neutron star matter across both low and high density regimes. A similar approach is adopted for quark stars. In particular, we construct a comprehensive dataset of equations of state using both the Massachusetts Institute of Technology (MIT) bag model~\cite{Haensel2007NeutronS1,Glendenning-2000,schaffner-bielich_2020} and the Color–Flavor–Locked (CFL) model~\cite{Alford-2001a,Alford-2008,Rajagopal-2000,Alford-1999}, from which the corresponding mass–radius relations are derived. This dataset is then used to train our classification model.
In the second step, all these  datasets are used to generate the corresponding mass–radius (M–R) relations, spanning the broadest possible range of masses and radii. Subsequently, ML and deep learning (DL) techniques are applied to analyze these configurations.

We use two complementary approaches to evaluate and validate the robustness of our results:
We employ four different ML algorithms  Random Forest ~\cite{Breiman-2001-RandomForest}, XGBoost ~\cite{Chen-2016-XGBoost}, Decision Tree~\cite{Quinlan-1986-ID3,Breiman-1984-CART}, and Logistic Regression ~\cite{Cox-1958-logistic} using the complete set of input features, and assess their performance in terms of accuracy and AUC. In addition, we perform an exhaustive search over all possible feature combinations to investigate potential feature redundancies and determine whether similar predictive performance can be achieved using fewer features.
To further validate our findings, we conduct a feature ablation study using a neural network architecture, systematically removing selective features and evaluating their impact on the overall model performance.

The paper is organized as follows: In Section II, we present in detail the two models employed for constructing the equations of state of both hadronic and quark matter. In Section III, we describe the dataset preparation procedure, while in Section IV we analyze the methodologies applied, with particular emphasis on the use of ML and DL models. Section V is devoted to the presentation of the results and the corresponding discussion. In Section VI, we introduce the synthetic-data generation procedure used to assess the robustness of the classifiers under uncertainties in  $M$, $R$, and $\Lambda$. The corresponding machine and deep learning results are presented and discussed in Section VII. The main conclusions of the present study are summarized in Section VIII with concluding remarks in Section IX. Finally, the Appendix includes a set of useful tables and graphical representations.
\section{The theoretical framework}
In this section, we present the fundamental assumptions and the underlying nuclear physics employed in the construction of the equations of state for neutron stars and quark stars. We also provide a brief overview of the differential equations governing the structure and macroscopic properties of compact objects, namely the Tolman–Oppenheimer–Volkoff (TOV) equations. Finally, we discuss one of the key properties characterizing the compact objects considered in this study, namely their tidal deformability, and present its theoretical formulation.

\subsection{Equations of state of neutron stars}
In the present analysis, beyond incorporating a diverse set of equations of state, it is essential to ensure extensive coverage of the mass-radius parameter space. Such coverage enables the construction of a sufficiently large dataset and enhances the generalizability of the models to be developed.  In this study, we assume the polytropic region extends over the interval [$\rho_0$, $7.5 \rho_0$], where $\rho_0=0.16 \ {\rm fm}^{-3}$ is the nuclear saturation density, which will also be denoted as $\rho_{\rm sat}$. We divide the region into four segments, with equal lengths in the logarithm of mass density $\rho$. For each segment, we allow four possible choices of $\Gamma_i$, namely $\{1,2,3,4\}$. We also employ HLPS-2 and HLPS-3, as the "main" EoSs before the transition to polytropic behavior~\cite{Hebeler-2013}.
To this end, a systematic strategy is adopted involving the artificial generation of a large ensemble of distinct equations of state. Within this framework, polytropic equations of state are employed as a practical and efficient means of generating the required dataset. In particular, we adopt the method introduced in Ref.~\cite{Read-2009} and subsequently applied in Refs.~\cite{raithel2016neutron,Ozel-2009}. Within this framework, the mass–density interval bounded by $\rho_{\min}$ and $\rho_{\max}$ is selected and partitioned into $n$ segments. The EoS is then parametrized by means of $n$ piecewise polytropic relations. Denoting the mass density and pressure at the boundaries of the $i$-th polytropic segment by $\rho_i$ and $P_i$, respectively, each segment is described by \cite{Read-2009}
\begin{equation}\label{press_poly}
\begin{matrix}
    P=K_i\rho^{\Gamma_i} & (\rho_{i-1}\leq\rho\leq\rho_i)
\end{matrix}
\end{equation}
where the constant $K_i$ is determined from the values of the pressure and mass density at the preceding fiducial point as follows
\begin{equation}\label{K-poly}
    K_i=\frac{P_{i-1}}{\rho^{\Gamma_i}_{i-1}}=\frac{P_i}{\rho^{\Gamma_i}_i}
\end{equation}
and the polytropic index of the segment $\Gamma_i$, is given by
\begin{equation}\label{Gamma-poly}
    \Gamma_i=\frac{\log_{10}(P_i/P_{i-1})}{\log_{10}(\rho_i/\rho_{i-1})}
\end{equation}
The value of $\Gamma_i$ within each segment is typically selected arbitrarily. The corresponding expressions for the polytropic equations of state are then obtained by integrating the equation
\begin{equation}
d\left(\frac{\epsilon}{\rho}  \right)=-Pd\left(\frac{1}{\rho}  \right)   
\label{dE-p} 
\end{equation}
which for the case $\Gamma\neq1$ leads to 
\begin{equation}
\label{polyEoSGn1form}
    \epsilon(\rho)=(1+a)\rho c^2+\frac{K}{\Gamma-1}\rho^{\Gamma}
\end{equation}
where $\alpha$ denotes an integration constant, the value of which is determined by enforcing continuity of the EoS across successive mass–density segments at the corresponding boundary points. Under this constraint, Eq.~(\ref{polyEoSGn1form}) assumes the form
\begin{eqnarray}\label{poly_EoS_Gn1_gen_form}
&&\epsilon(\rho)=\left[\frac{\epsilon(\rho_{i-1})}{\rho_{i-1}}-\frac{P_{i-1}}{\rho_{i-1}(\Gamma_i -1)}\right]\rho + \frac{K_i}{\Gamma_i-1}\rho^{\Gamma_i}, \nonumber \\
&& \text{ }\text{ } (\rho_{i-1}\leq\rho\leq\rho_i)
\end{eqnarray}
where $K_i$ and $\Gamma_i$ are calculated  
from Eqs.~(\ref{K-poly}) and ~(\ref{Gamma-poly}), respectively.
In the same way, integrating Eq.~(\ref{dE-p}), for the case  $\Gamma=1$, the energy density reads \cite{raithel2016neutron}
\begin{eqnarray}\label{poly_EoS_G1_gen_form}
&&\epsilon(\rho)=\frac{\epsilon(\rho_{i-1})}{\rho_{i-1}}\rho + K_i\ln\left(\frac{1}{\rho_{i-1}}\right)\rho-K_i\left(\frac{1}{\rho}\right)\rho, \nonumber\\
&&\text{ }\text{ } (\rho_{i-1}\leq\rho\leq\rho_i)
\end{eqnarray}

We can rewrite 
Eqs.~(\ref{poly_EoS_Gn1_gen_form}) and ~(\ref{poly_EoS_G1_gen_form}) with the energy density being a function of pressure $\epsilon(P)$. To do so, we use the polytropic relation between pressure and mass density from Eq.~(\ref{press_poly}). For $\Gamma\neq1$, we have
\begin{eqnarray}\label{poly_EoS_Gn1_gen_form2}
&&\epsilon(P)=\left[\frac{\epsilon(\rho_{i-1})}{\rho_{i-1}}-\frac{P_{i-1}}{\rho_{i-1}(\Gamma_i -1)}\right]\left(\frac{P}{K_i}\right)^{\Gamma_i^{-1}}+\frac{P}{\Gamma_i-1}, \nonumber\\
&&\text{ }\text{ } (P_{i-1}\leq P\leq P_i)
\end{eqnarray}
and for $\Gamma=1$
\begin{eqnarray}\label{poly_EoS_G1_gen_form2}
&&\epsilon(P)=\frac{\epsilon(\rho_{i-1})}{\rho_{i-1}}\frac{P}{K_i}+\ln\left(\frac{1}{\rho_{i-1}}\right)P-\ln\left(\frac{K_i}{P}\right)P, \nonumber \\
&&\text{ }\text{ } (P_{i-1}\leq P\leq P_i)
\end{eqnarray}
It is important to emphasize that, in the present study, the equations of state are constructed to rigorously satisfy the causality condition, which, as dictated by special relativity, imposes an upper bound on the pressure gradient
\begin{equation}\label{caus_limit}
    \frac{dP}{d\epsilon}\equiv\left(\frac{c_s}{c}\right)^2\leq1
\end{equation}
commonly referred to as the causality limit, where $c_s/c$ denotes the speed of sound $c_s$
expressed in units of the speed of light $c$. 
Accordingly, all equations of state considered herein are constructed to rigorously satisfy this constraint under all conditions. This requirement enhances the physical consistency and reliability of the artificially generated equations of state, while ensuring their proper behavior in the high-density regime.

It is possible, however, that some of the mock polytropic EoSs, we construct violate the condition of Eq.~(\ref{caus_limit}) after a certain value of mass density $\rho_{tr}$ (or equivalently pressure $P_{tr}$). In those cases, we assume the transition from the polytropic parametrized EoS to an EoS of linear behavior, at pressure $P_{tr}$ and beyond. A Maxwell construction is well-suited to describe this kind of transition \cite{laskos2025speed}
\begin{equation}\label{Maxwell_construct}
    \epsilon(P) = \begin{cases}
    \epsilon_{\rm Hadronic}(P),&P\leq P_{tr} \\
    \epsilon(P_{tr}) +\Delta\epsilon+(c_s/c)^{-2}(P-P_{tr}), &P>P_{tr}
    \end{cases}
\end{equation}
where $\epsilon_{\rm Hadronic}(P)$ in the first line of Eq.~(\ref{Maxwell_construct}) stands for the hadronic phase before the transition, governed by a continuous EoS (polytropic or other), and the second line refers to the maximally stiff high density phase. Note that there is no mixed phase region (as in Gibbs construction) and that the two phases co-exist only at the phase transition pressure $P_{tr}$. Lastly, the term $\Delta\epsilon$ establishes the discontinuity of the energy density and hence the discontinuity of the total EoS at the transition density  $\rho_{tr}$ (or pressure $P_{tr}$). However, in our case we consider a continuous total EoS, so the term $\Delta \epsilon$ vanishes. Furthermore, we impose the extreme causality condition given by Eq.~(\ref{caus_limit}), namely, we set the slope of the linear EoS to satisfy
$(c_s/c)^{-2}=1$.

Finally, in the present study, we employ as a test set a representative sample of equations of state derived from both microscopic and phenomenological approaches, which have been widely used in the neutron-star literature (see Ref.~\cite{Koliogiannis-2020} and references therein). The primary objective is to assess the accuracy of the proposed method in reproducing these equations of state, under the assumption that the corresponding mass–radius (M–R) relations are known. Care has been taken to ensure that the selected models span a broad region of the M–R parameter space, thereby capturing, to the greatest extent possible, the diversity of existing theoretical predictions. A common characteristic of all selected equations of state is that they predict a maximum mass exceeding two solar masses.

\subsection{Equations of state of quark stars} 
For the study of quark stars and the construction of the corresponding equations of state, we employ two theoretical frameworks: (a) the MIT bag model and (b) the CFL model.
\\
{\bf a. The MIT Bag Model:}
The simplest phenomenological framework for describing quark-star matter is the well-established bag model~\cite{Haensel2007NeutronS1,schaffner-bielich_2020}. In particular, the MIT bag model treats quark matter as a gas of free, relativistic quarks confined within a finite region (“bag”), which is stabilized by a vacuum pressure $B$. This parameter, assumed to be constant, is commonly referred to as the MIT bag constant~\cite{schaffner-bielich_2020}. Within this framework, the resulting EoS  takes the form of a linear relation between pressure and energy density, and can be expressed as

\begin{equation}
P=\frac{1}{3}\left(\epsilon-4B\right)
\label{MIT_EOS_press}
\end{equation}
Despite its simplicity, stemming from the fact that the EoS  is governed by a single parameter into which all interaction effects are effectively absorbed, the model provides an adequate description of the fundamental properties of quark stars, provided an appropriate choice of the parameter $B$ is made. In the present work, it is employed primarily as a benchmark for comparison with more sophisticated models, such as the CFL model discussed below.
\\
{\bf b. The Color-Flavor Locked  Model:}
A substantial body of work has been devoted to Witten’s proposal concerning the true ground state of strongly interacting matter~\cite{Witten-1984}, an idea that was originally anticipated in the seminal work of Bodmer~\cite{Bodmer-1971}. According to this hypothesis, quark matter composed of u, d, and s quarks, commonly referred to as strange matter, may exhibit an energy per baryon lower than that of both nuclear matter and two-flavor (u–d) quark matter.
In conjunction with theoretical expectations for the appearance of deconfined quark matter at supranuclear densities~\cite{Ivanenko-1965,Itoh-1970,Collins-1975,Weber-2005,schaffner-bielich_2020}, this conjecture has stimulated extensive investigations into the possible existence of exotic compact objects known as strange stars~\cite{schaffner-bielich_2020,Glendenning-2000,Alock-1986,Hanesel-1986}. Owing to their distinctive composition, such objects are predicted to sustain configurations with arbitrarily small masses and radii~\cite{schaffner-bielich_2020}.
At asymptotically high densities quark masses become negligible relative to the chemical potential. Under such conditions, quark matter is expected to enter a superfluid phase in which quarks of all flavors and colors form Cooper pairs with a common Fermi momentum. This phase is known as the CFL phase~\cite{Alford-2001a,Alford-2008,Rajagopal-2000,Alford-1999,oikonomou2023color,flores2017constraining}.
Self-bound compact stars composed entirely of quark matter, referred to as strange stars, may arise over a broad range of parameters within the MIT bag model EoS~\cite{lugones2002color}. Studies of their structure further indicate that color superconductivity can significantly modify their mass–radius relation, allowing for substantially larger maximum masses~\cite{lugones2003high,horvath2004self}.

In particular, the  EoS  for CFL quark matter can be formulated within the general framework of the MIT bag model. The pressure and energy density are given, in order of $\Delta^2$ ($\Delta$ is the gap parameter representing the contribution of color superconductivity \cite{oikonomou2023color}) and $m_s^2$ (with $m_s$ the mass of the strange quark), as follows \cite{flores2017constraining}
\begin{eqnarray}\label{CFL_EOS_enrg2}
    \epsilon &=& 3P+4B-\frac{9\alpha\mu^2}{(\hbar c)^3\pi^2}, \nonumber \\  \mu^2&=&-3\alpha+\left(\frac{4}{3}\pi^2(B+P)(\hbar c)^3+9\alpha^2\right)^{1/2}  
\end{eqnarray}
where
\begin{equation}\label{alpha_cfl}
    \alpha=-\frac{m_s^2}{6}+\frac{2\Delta^2}{3}
\end{equation}
The absolute stability of the CFL quark matter requires the energy per baryon to be less than the neutron mass $m_n$ at vanishing pressure ($P=0$) and temperature ($T=0$). Thus, the following condition must be satisfied \cite{lugones2002color}
\begin{equation}\label{CFL_stable_baryonenrg}
    \frac{\epsilon}{n_B}\bigg|_{P=0}=3\mu\leq m_n = 939 \text{ MeV}
\end{equation}
This result is derived directly from the shared Fermi momentum among the three quark flavors in CFL matter and is valid at $T=0$ without any approximation. Since this condition must be fulfilled at  the vanishing-pressure point, using the second relation of Eq.~(\ref{CFL_EOS_enrg2}), we get \cite{flores2017constraining}
\begin{equation}\label{bag_high_constraint}
    B<-\frac{1}{(\hbar c)^3}\left(\frac{m_s^2m_n^2}{12\pi^2}-\frac{\Delta^2m_n^2}{3\pi^2}-\frac{m_n^4}{108\pi^2}\right)
\end{equation}
This equation defines a region in the $m_s$-$B$ plane in which
the energy per baryon is smaller than $m_n$ for a given $\Delta$. 

\subsection{The TOV equations }
The mechanical equilibrium of stellar matter is governed by a coupled system of two differential equations: the well-known TOV equations, together with the EoS  of the fluid, ${\cal \epsilon} = {\cal \epsilon}(P)$. This system can be expressed as follows~\cite{Shapiro:1983du,Haensel2007NeutronS1,Glendenning-2000,schaffner-bielich_2020}
\begin{eqnarray}
\frac{dP(r)}{dr}&=&-\frac{G{\cal \epsilon}(r) M(r)}{c^2r^2}\left(1+\frac{P(r)}{{\cal \epsilon}(r)}\right) \nonumber \\
&\times&
 \left(1+\frac{4\pi P(r) r^3}{M(r)c^2}\right) \left(1-\frac{2GM(r)}{c^2r}\right)^{-1},
\label{TOV-1}
\end{eqnarray}
\begin{equation}
\frac{dM(r)}{dr}=\frac{4\pi r^2}{c^2}{\cal \epsilon}(r).
\label{TOV-2}
\end{equation}
The solution of the coupled differential equations (\ref{TOV-1}) and (\ref{TOV-2}) for $P(r)$ and $M(r)$ necessitates their numerical integration from the stellar center ($r = 0$) outward to the radius $r = R$, where the pressure effectively vanishes. At this point, the corresponding values of the stellar radius and gravitational mass are determined.

\subsection{Tidal Deformability}

One of the most significant sources for the terrestrial gravitational-wave detectors is gravitational waves from inspiraling binary NS systems before their merger~\cite{Postnikov-2010,Baiotti-2019,Flanagan-08,Hinderer-08,Damour-09,Hinderer-10,Koliogiannis-2021}. The component masses of these binary systems can be measured. Additionally, during the last orbits before the merger, the tidal effects that are present can also be measured~\cite{Flanagan-08}.

The dimensionless parameter that describes the response of a NS to the induced tidal field is called the tidal Love number $k_2$. This parameter depends on the NS structure (i.e. the mass of the neutron star, and the EoS). Specifically, the tidal Love number $k_2$ relates the induced quadrupole moment to applied tidal field  $Q_{ij}$ and the applied tidal field $E_{ij}$~\cite{Flanagan-08}, given below

\begin{equation}
Q_{ij}=-\frac{2}{3}k_2\frac{R^5}{G}E_{ij}\equiv- \lambda E_{ij},
\label{Love-1}
\end{equation}
where $R$ is the NS radius and $\lambda=2R^5k_2/3G$ is a key quantity, which is called tidal deformability. The tidal Love number $k_2$ is given by \cite{Flanagan-08,Hinderer-08}
\begin{eqnarray}
k_2&=&\frac{8C^5}{5}\left(1-2C\right)^2\left[2-y_R+(y_R-1)2C \right]
\nonumber \\
&\times&
\left[\frac{}{} 2C \left(6  -3y_R+3C (5y_R-8)\right) \right. \nonumber \\
&+& 4C^3 \left.  \left(13-11y_R+C(3y_R-2)+2C^2(1+y_R)\right)\frac{}{} \right.\nonumber \\
&+& \left. 3\left(1-2C \right)^2\left[2-y_R+2C(y_R-1)\right] {\rm ln}\left(1-2C\right)\right]^{-1},
\label{k2-def}
\end{eqnarray}
where $C$ is the compactness parameter $C=GM/Rc^2$ while  
the quantity $y_R$ is determined by solving the following differential equation
\begin{equation}
r\frac{dy(r)}{dr}+y^2(r)+y(r)F(r)+r^2Q(r)=0, 
\label{D-y-1}
\end{equation}
with the initial condition $ y(0)=2$~\cite{Hinderer-10}. $F(r)$ and $Q(r)$ are functionals of $\epsilon(r)$, $P(r)$ and $M(r)$  defined as~\cite{Postnikov-2010,Hinderer-10}
\begin{equation}
F(r)=\left[ 1- \frac{4\pi r^2 G}{c^4}\left(\epsilon(r)-P(r) \right)\right]\left(1-\frac{2M(r)G}{rc^2}  \right)^{-1},
\label{Fr-1}
\end{equation}
and 
\begin{eqnarray}
r^2Q(r)&=&\frac{4\pi r^2 G}{c^4} \left[5\epsilon(r)+9P(r)+\frac{\epsilon(r)+P(r)}{\partial P(r)/\partial\epsilon (r)}\right] \nonumber \\
&\times&
\left(1-\frac{2M(r)G}{rc^2}  \right)^{-1}
-6\left(1-\frac{2M(r)G}{rc^2}  \right)^{-1}
\nonumber \\
&-&\frac{4M^2(r)G^2}{r^2c^4}\left(1+\frac{4\pi r^3 P(r)}{M(r)c^2}   \right)^2 \nonumber \\
&\times& \left(1-\frac{2M(r)G}{rc^2}  \right)^{-2}.
\label{Qr-1}
\end{eqnarray}
Eq.~(\ref{D-y-1})  must be integrated self-consistently, together with the  TOV equations using the boundary conditions $y(0)=2$, $P(0)=P_c$ and $M(0)=0$~\cite{Postnikov-2010,Hinderer-08}. The numerical solution of these equations provides the mass $M$, the radius $R$ of the NS, and the value of $y_R=y(R)$. The latter parameter along with the quantity $\beta$ are the  basic ingredients  of the tidal Love number $k_2$.

In addition, an important and well-measured quantity by the gravitational-wave detectors, which can be treated as a tool to impose constraints on the EoS, is the dimensionless tidal deformability $\Lambda$, defined as 
\begin{equation}
    \Lambda=\frac{2}{3}k_2 \left(\frac{c^2R}{GM}\right)^5
\label{dimens-tidal}    
\end{equation}
We note that $\Lambda$ is sensitive to the neutron star radius, hence can provide information for the low-density part of the EoS, which is also related to the structure and properties of finite nuclei.

\section{Dataset Preparation}
This study employs ML techniques to classify compact stars, specifically neutron stars and quark stars. The dataset used consists of  mass, radius, tidal deformability, Love number ($k_2$) and central pressure as features, with the goal of distinguishing between the two classes. It comprises 37528 samples, with two classes: neutron stars and quark stars belonging to a total of 3952 EoSs, nearly half of them of hadronic type (2008) and the other half quark type (1944). For each EoS, we calculate different values for $M,R,k_2, \Lambda, P_c$ that span the entire M-R line.
\subsection{Training strategy}
We additionally applied stratified group k-fold cross-validation using the corresponding  function of the scikit-learn library~\cite{Pedregosa-2011} in order to perform a more rigorous evaluation of the classification models. In contrast to a simple random split, where individual samples are randomly distributed between training and testing sets, stratified group k-fold cross validation ensures that all samples belonging to the same group, defined in the dataset by an EoS identifier, are assigned exclusively either to the training or to the validation fold. This prevents potential information leakage between datasets that could otherwise lead to artificially inflated performance estimates. At the same time, the stratified nature of the procedure preserves, as much as possible, the class distribution across the folds, which is particularly important for classification tasks with imbalanced classes. Using this strategy, the dataset was partitioned into independent folds based on the group EoS identifiers, and each model was trained and evaluated across all folds. Cross-validation therefore provides a more reliable estimate of the evaluation metrics, such as AUC and accuracy, by reducing the risk of overfitting and minimizing the dependence of the results on a single train–test split.

\subsection{Feature Relationships}
To explore the relationships between the input features, we constructed a Pearson correlation matrix. This matrix quantifies the linear relationships between the features, providing insight into which features are most strongly correlated. We observed that the radius $R$ and the Love number ($k_2$) exhibited the strongest correlations with a moderate correlation coefficient of 0.64. In all other cases, the correlations were weaker.

\begin{figure}[t]
\centering
\includegraphics[width=0.4\textwidth]
{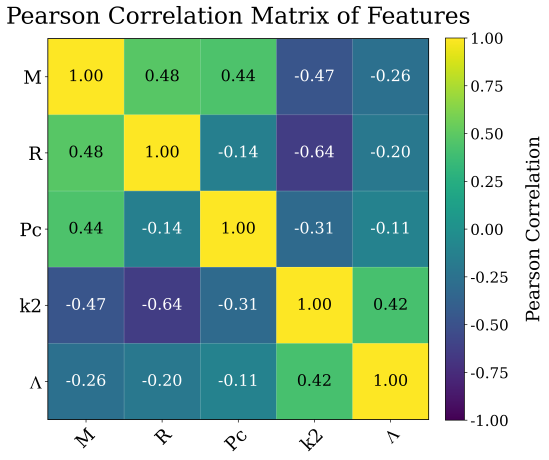}
\caption{Pearson correlation matrix showing the linear correlations between the selected features.}
\label{fig:correlation_matrix}
\end{figure}

\section{Machine and Deep Learning Models Used for the Baseline Dataset}

We used two different strategies in order to assess and validate our results:

1. We employed four different ML models Random Forest~\cite{Breiman-2001-RandomForest}, XGBoost~\cite{Chen-2016-XGBoost}, Decision Tree~\cite{Quinlan-1986-ID3,Breiman-1984-CART}, and Logistic Regression ~\cite{Cox-1958-logistic} to evaluate the classification performance in terms of accuracy and AUC using the full feature set. Furthermore, we performed an exhaustive feature-subset search to investigate potential feature degeneracies, namely whether comparable performance could be achieved using fewer features.

2. To further validate our findings, we applied feature ablation analysis in a neural network by systematically removing features and evaluating the impact on model performance.

\subsection{Machine Learning Models}

Four classifiers were trained and evaluated to classify neutron and quark stars: Random Forest, XGBoost, Decision Tree and Logistic Regression. These classifiers were optimized using grid search with four-fold cross-validation. The models were evaluated using several metrics, including accuracy, precision, recall, F1-score and AUC. 

\begin{figure*}[ht]
\centering
\includegraphics[width=0.75\textwidth]{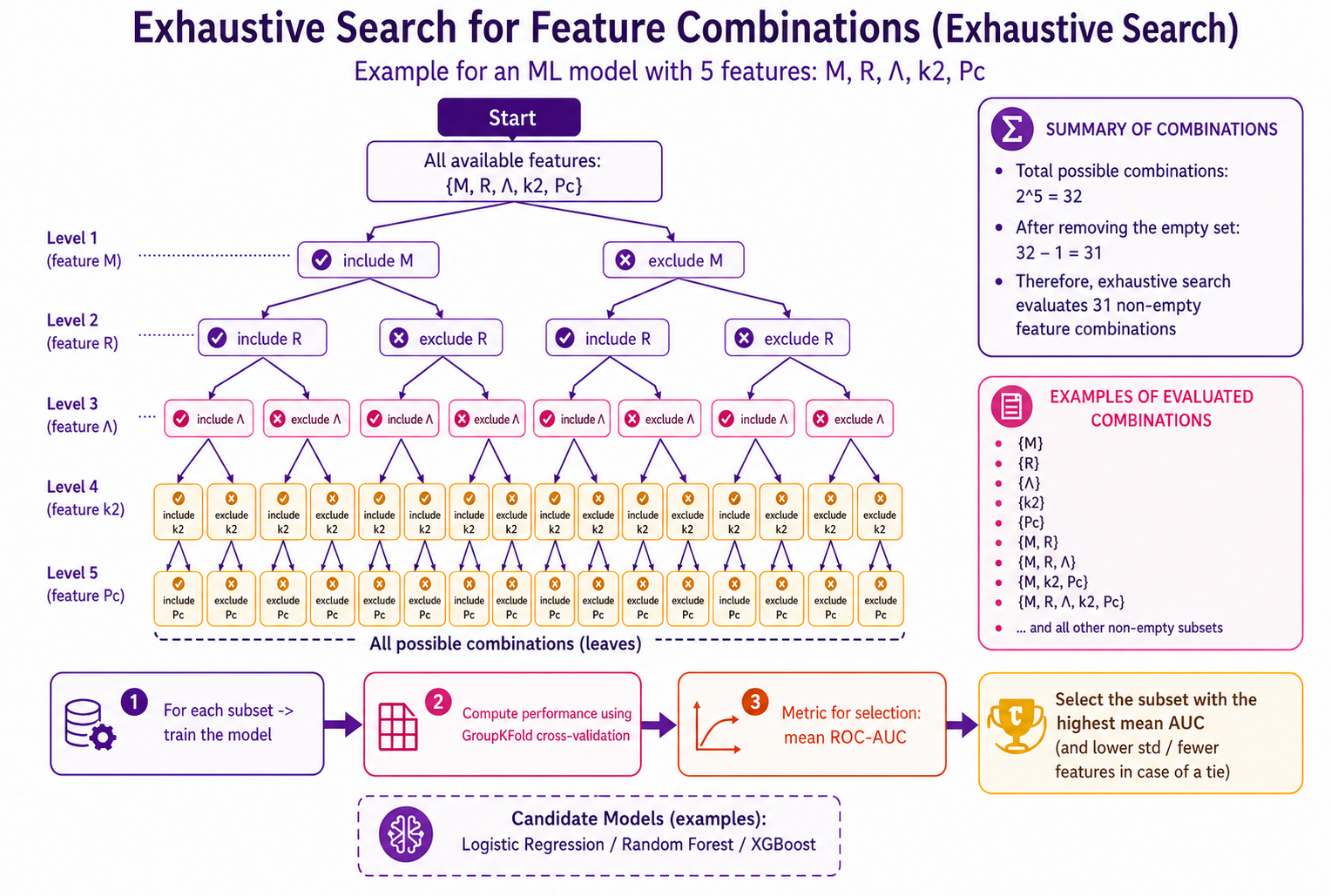}
\caption{Schematic representation of the exhaustive feature-subset search. Starting from the full set of five physical quantities, all possible non-empty subsets are generated and evaluated using stratified group k-fold cross-validation. The optimal subsets are selected according to the mean ROC-AUC score, while the standard deviation is used as an additional measure of stability.}
\label{fig:exhaustive_search}
\end{figure*}

To investigate the relative importance of the physical input features, an exhaustive feature-subset search was performed. Since the initial feature set consists of five physical quantities, namely the mass $M$, radius $R$, tidal deformability $\Lambda$, Love number $k_2$, and central pressure $P_c$, a total of
$
2^5 - 1 = 31
$
feature subsets were examined for each classifier.
The mean ROC-AUC score was used as the main selection criterion, while the standard deviation across folds was also recorded in order to assess the stability of each feature combination. 
For each classifier, the three highest-ranked feature subsets were retained. 
In addition, the subset $\{M,R,\Lambda\}$ was explicitly included in the comparison, since these quantities correspond to observables that are more directly connected with astrophysical measurements. 
The ranking position of this subset among all 31 possible combinations is also reported.
A schematic representation of the exhaustive-search procedure is shown in Figure~\ref{fig:exhaustive_search}.

\subsection{Neural Network Training}
\begin{figure*}[ht]
\includegraphics[width=0.7\linewidth]
{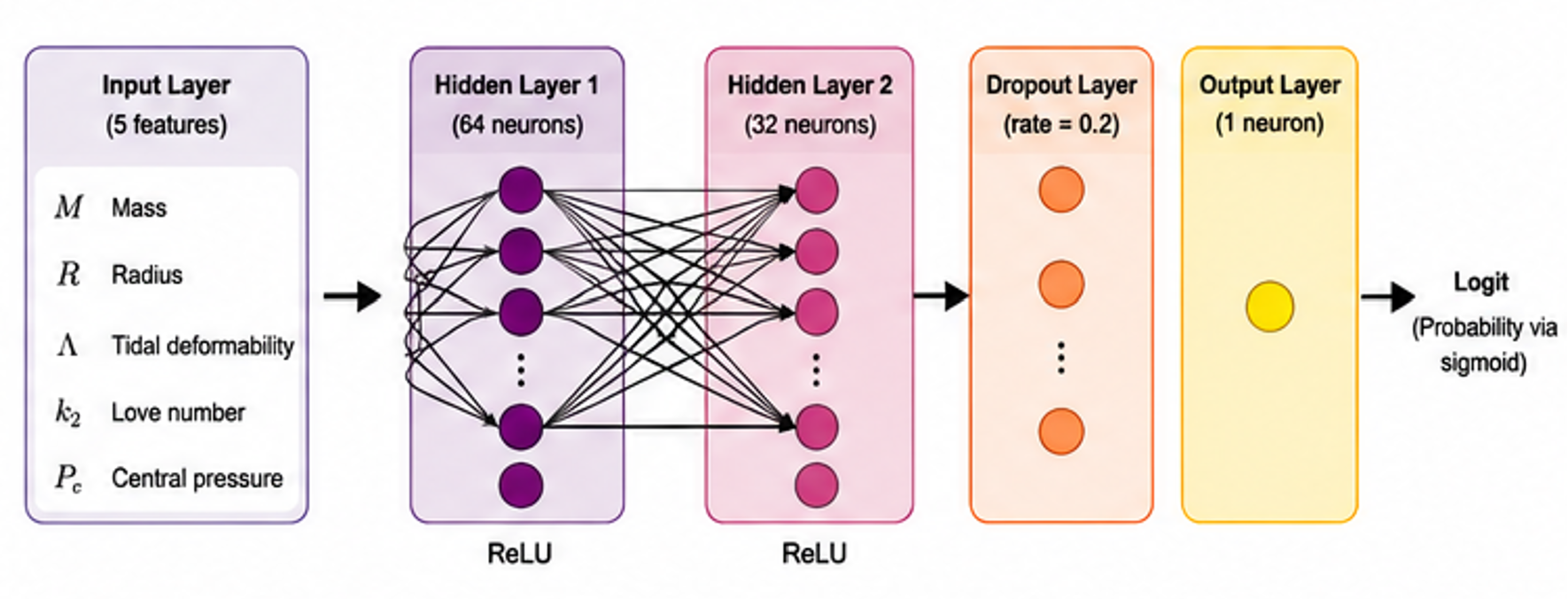}
\caption{Schematic architecture of the feedforward neural network used for the classification of compact stars. The model receives as input five physical quantities ($M$, $R$, $\Lambda$, $k_2$, and $P_c$), followed by two fully connected hidden layers with 64 and 32 neurons, respectively. ReLU activation functions are applied to the hidden layers, while a dropout layer with rate 0.2 is included to avoid overfitting. The final output layer consists of a single neuron that produces the classification logit for the binary decision between neutron stars and quark stars.}
    \label{fig:nn_architecture}
\end{figure*}

To complement the classical ML classifiers, we developed a feedforward neural-network model \cite{Rumelhart-1986-Backprop} for the binary classification of compact stars into neutron stars and quark stars. The network was constructed using the same dataset and the five physical input features considered throughout this work, namely the mass $M$, radius $R$, tidal deformability $\Lambda$, Love number $k_2$, and central pressure $P_c$. 
The adopted architecture consists of an input layer with five neurons, followed by two fully connected hidden layers with 64 and 32 neurons, respectively. Each hidden layer employs the Rectified Linear Unit (ReLU) activation function. In order to reduce the risk of overfitting and improve generalization, a dropout layer with rate 0.2 was applied after the second hidden layer. The output layer consists of a single neuron corresponding to the binary classification task. During training, the network parameters were optimized using the Adam optimizer together with the binary cross-entropy loss with logits (BCEWithLogitsLoss). A schematic representation of the neural-network architecture is presented in Figure~\ref{fig:nn_architecture}.

\section{Results and Discussion for the Baseline Dataset}
In this section, we present the results of the ML models
and neural-network classification, evaluating their performance based on
accuracy, precision, recall, F1-score, and AUC. We also discuss the
results of the confusion matrices and ROC curves.

\subsection{Performance of Machine Learning Models}
To evaluate the performance of the ML classifiers, we used
Random Forest, XGBoost, Decision Tree, and Logistic Regression.
Hereafter, the abbreviations RF, XGB, DT, and LR denote Random Forest,
XGBoost, Decision Tree, and Logistic Regression, respectively. The classifiers were evaluated using accuracy, precision, recall,
F1-score, and AUC.

\begin{figure*}[ht]
\centering
\includegraphics[width=0.39\textwidth]{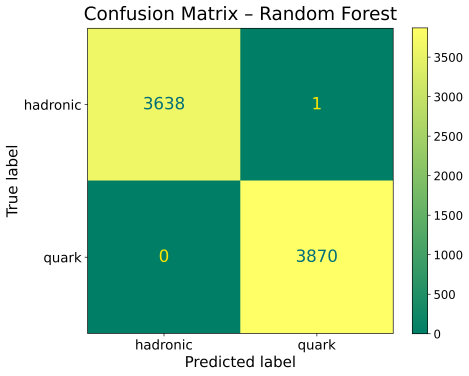}
\includegraphics[width=0.4\textwidth]{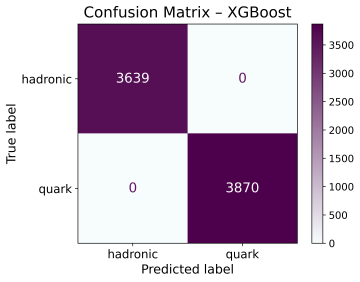} \includegraphics[width=0.4\textwidth]{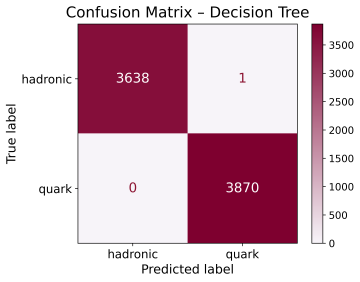}
\includegraphics[width=0.4\textwidth]{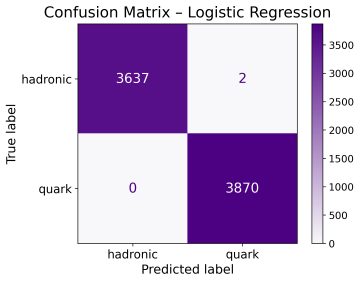}
\caption{Confusion matrices of the four ML  classifiers using the full five-feature set
$(M, R, \Lambda, k_2, P_c)$: Random Forest  (upper left), XGBoost  (upper right), Decision Tree  (lower left) and  Logistic Regression  (lower right).  }
  \label{ConfusionMatrix}
\end{figure*}
\begin{figure}[t]
\centering \includegraphics[width=0.43\textwidth]{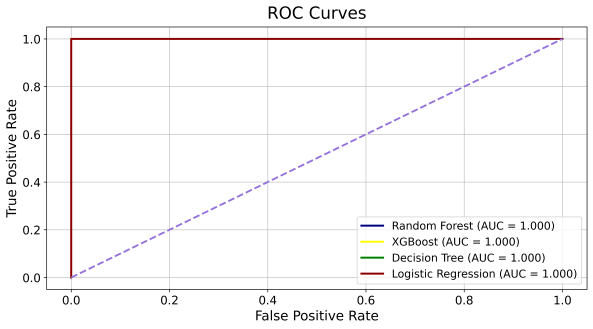}
\caption{ROC curves of the four ML classifiers using the full five-feature set
$(M, R, \Lambda, k_2, P_c)$.}
\label{fig:roc_curves}
\end{figure}

\begin{table}[H]
  \caption{Performance metrics of the ML classifiers evaluated using the stratified group k-fold cross-validation strategy.}
\centering
\begin{tabular}{|c|c|c|c|c|c|}
    \hline
    \textbf{Metrics} & \textbf{RF} &\textbf{XGB} & \textbf{DT} & \textbf{LR} \\
        \hline
        Accuracy  & 0.9999 & 1.0000 & 0.9999 & 0.9997 \\
        Precision & 0.9997 & 1.0000 & 0.9997 & 0.9995 \\
        Recall    & 1.0000 & 1.0000 & 1.0000 & 1.0000 \\
        F1-score  & 0.9999 & 1.0000 & 0.9999 & 0.9997 \\
        AUC-score & 1.0000 & 1.0000 & 0.9999 & 1.0000 \\
    \hline
  \end{tabular}
  \label{tab:ml_metrics}
\end{table}
Table~\ref{tab:ml_metrics} shows the performance metrics of these
classifiers, which achieved near-perfect classification results. The confusion matrices for each of the ML classifiers (Random Forest, XGBoost, Decision Tree and Logistic Regression) are shown in Figure~\ref{ConfusionMatrix}. 
The Receiver Operating Characteristic (ROC) curves for the four classifiers are presented in Figure \ref{fig:roc_curves}, with each classifier achieving an AUC of 1.0, demonstrating excellent discriminative capability of all four classifiers.
To complement the performance analysis based on the full five-feature set,
we next examine the results of the exhaustive feature-subset search described in Sec.~IV.A. For each classifier, all subsets of the five input features were ranked according to their mean ROC-AUC score across the cross-validation folds. The corresponding standard deviation was also reported as a measure of the stability of each subset across the folds.
The results show that subsets containing the Love number $k_2$ systematically appear among the highest-ranked combinations for all classifiers. 
In particular, the subset $\{M,k_2,P_c\}$ provides the best performance for both Random Forest and XGBoost, while closely related combinations involving $k_2$ dominate the highest-ranked subsets of Decision Tree and Logistic Regression as well. 
On the other hand, the observationally motivated subset $\{M,R,\Lambda\}$ performs less favorably in the exhaustive ranking, occupying the 14th, 16th, 18th, and 19th positions for Random Forest, XGBoost, Decision Tree, and Logistic Regression, respectively. 
This indicates that although mass, radius, and tidal deformability contain useful information, the inclusion of the Love number $k_2$ significantly improves the separability between neutron stars and quark stars.
\begin{table}[h!]
\caption{Feature-subset ranking obtained through exhaustive search. For each classifier, the three best-performing subsets are reported, together with the subset $\{M,R,\Lambda\}$ and its ranking position among the 31 possible non-empty feature combinations.}
\centering
\label{tab:exhaustive_search_results}
\begin{tabular}{llcccc}
\hline
Classifier & Feature Subset & Rank & Mean AUC & Std  \\
\hline
RF & $\{M,k_2,P_c\}$ & 1 & 1.00000000 & 0.00000000 \\
RF & $\{M, \Lambda,k_2\}$ & 2 & 1.00000000 & 0.00000000 \\
RF & $\{M,R,k_2,P_c\}$ & 3 & 1.00000000 & 0.00000000 \\
RF & $\{M,R,\Lambda\}$ & 14 & 0.99998762 & 0.00000880 \\
\hline
XGB & $\{M,k_2,P_c\}$ & 1 & 1.00000000 & 0.00000000 \\
XGB& $\{M,R,k_2,P_c\}$ & 2 & 1.00000000 & 0.00000000  \\
XGB & $\{M, \Lambda,k_2,P_c\}$ & 3 & 1.00000000 & 0.00000000  \\
XGB & $\{M,R,\Lambda\}$ & 16 & 0.99647077 & 0.00083398\\
\hline
DT & $\{\Lambda,k_2,P_c\}$ & 1 & 0.99986654 & 0.00013364 \\
DT & $\{\Lambda,k_2\}$ & 2 & 0.99983440 & 0.00017369 \\
DT & $\{M,k_2,\Lambda\}$ & 3 & 0.99980235 & 0.00015045  \\
DT & $\{M,R,\Lambda\}$ & 18 & 0.98323151 & 0.00075005 \\
\hline
LR & $\{M,R,k_2\}$ & 1 & 1.00000000 & 0.00000000 \\
LR & $\{M,R,P_c,k_2\}$ & 2 & 1.00000000 & 0.00000000 \\
LR & $\{M,R,k_2,\Lambda\}$ & 3 & 1.00000000 & 0.00000000 \\
LR & $\{M,R,\Lambda\}$ & 19 & 0.84836675 & 0.00179339 \\
\hline
\end{tabular}
\end{table}

\subsection{Neural Network Performance}

\begin{figure}[t]
\centering
\includegraphics[width=0.9\linewidth]{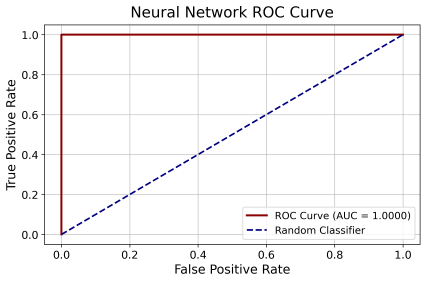}
\caption{ROC curve of the neural-network classifier using the full five-feature set
$(M, R, \Lambda, k_2, P_c)$.}
    \label{fig:roc_neural_network}
\end{figure}

In addition to the classical ML models, a neural network with two hidden layers was trained using the same dataset $\{M,R,\Lambda,k_2,P_c\}$. The neural network achieved 100\% accuracy, 100\% precision, 100\% recall, an F1-score of 1.0, and an AUC-score of 1.0, indicating performance comparable to that of the traditional ML models.
To further evaluate the discriminative capability of the neural-network classifier, the ROC curve was also examined. The ROC curve provides a threshold-independent assessment of classification performance by illustrating the trade-off between the true positive rate and the false positive rate across different decision thresholds (see  Figure~\ref{fig:roc_neural_network}).

\begin{figure}[t]
\centering
\includegraphics[width=0.44\textwidth]{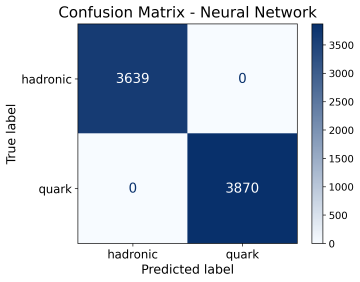}
\caption{Confusion matrix of the neural-network classifier using the full five-feature set
$(M, R, \Lambda, k_2, P_c)$.}
\label{fig:nn_confusion_matrix}
\end{figure}

\begin{table}[H]
  \caption{Performance metrics for the neural network model.}
\centering
\begin{tabular}{|c|c|}
    \hline
    \textbf{Results of neural network} & \textbf{Value} \\
    \hline
    Accuracy & 1.00 \\
    \hline
    Precision & 1.00 \\
    \hline
    Recall & 1.00 \\
    \hline
    F1-Score & 1.00 \\
    \hline
    AUC-Score & 1.00 \\
    \hline
  \end{tabular}
\label{tab:neural_network_performance}
\end{table}
To provide an additional assessment of the predictive capability of the neural-network classifier, the confusion matrix for the test set is presented in Figure~\ref{fig:nn_confusion_matrix}. The matrix demonstrates perfect classification performance, with all hadronic and quark stars correctly identified and no misclassifications observed.
The absence of off-diagonal elements indicates that the neural network achieved complete separability between the two stellar classes within the examined dataset. These results are fully consistent with the ROC analysis and the corresponding AUC-score of 1.00, further confirming the robustness and reliability of the neural-network approach for compact-star classification.

\begin{figure}[t]
\centering
\includegraphics[width=0.46\textwidth]{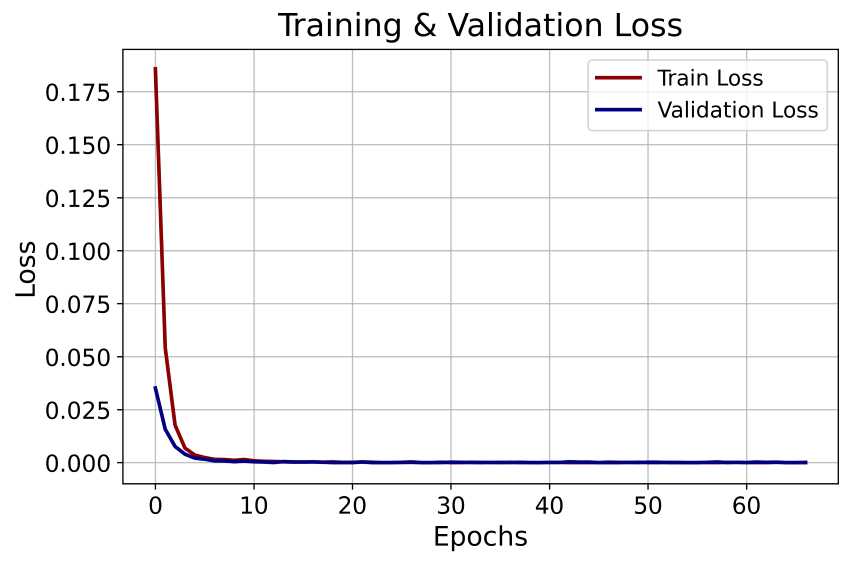}
\caption{Training and validation loss curves of the neural network across epochs. Both curves converge rapidly to low values, indicating fast and stable training without overfitting.}
\label{fig:training_loss}
\end{figure}

As shown in Figure~\ref{fig:training_loss}, both the training and validation loss decrease rapidly during the first epochs and converge smoothly to very low values. The close agreement between the two curves indicates that the neural network learns stable and generalizable patterns, without signs of significant overfitting. This behavior confirms the efficiency of the training process and supports the reliability of the selected physical features for distinguishing between neutron stars and quark stars.

To assess the discriminative capability of the most relevant physical quantities identified by the exhaustive feature-subset search, the neural-network classifier was retrained using the two highest-ranked feature combinations obtained from the ML analysis. In particular, the subsets $(M, P_c, k_2)$ and $(M, k_2, \Lambda)$ were selected, since these combinations consistently achieved the best classification performance across the examined classifiers.

The objective of this analysis is to examine whether the neural network can preserve its predictive performance using a reduced number of physically meaningful variables, while also assessing the relative importance of the selected quantities for the separation between hadronic and quark stars.

Table~\ref{tab:nn_feature_triplets} summarizes the neural-network performance obtained for the two selected feature triplets.

\begin{table}[H]
 \caption{Performance of the neural-network classifier for the full five-feature set and the two selected feature triplets identified through the exhaustive feature-subset search. Acc, Prec, Rec, and F1 denote accuracy, precision, recall, and F1-score, respectively, while AUC denotes the area under the ROC curve.}
  \centering
  \begin{tabular}{|c|c|c|c|c|c|}
    \hline
    \textbf{Features} & \textbf{Acc} & \textbf{Prec} & \textbf{Rec} & \textbf{F1} & \textbf{AUC} \\
    \hline
    $M,R,P_c,k_2,\Lambda$ & 1.0000 & 1.0000 & 1.0000 & 1.0000 & 1.0000 \\
    \hline
    $M, P_c, k_2$ & 1.0000 & 1.0000 & 1.0000 & 1.0000 & 1.0000 \\
    \hline
    $M, k_2, \Lambda$ & 0.9647 & 0.9811 & 0.9499 & 0.9652 & 0.9971 \\
    \hline
  \end{tabular}
\label{tab:nn_feature_triplets}
\end{table}

As shown in Table~\ref{tab:nn_feature_triplets}, the reduced triplet $(M,P_c,k_2)$ preserves perfect classification performance, yielding accuracy, precision, recall, F1-score, and AUC all equal to 1.0000. This demonstrates that the combination of stellar mass, central pressure, and Love number retains essentially all the discriminative information required by the neural network. 
The triplet $(M,k_2,\Lambda)$ leads to a lower, although still very strong, classification performance. The accuracy decreases to 0.9647, while the F1-score remains high at 0.9652 and the AUC reaches 0.9971. The high AUC value indicates that the model still has excellent discriminative capability.

\section{Creating synthetic data to assess classifier robustness}

The main purpose of the study is to construct models in order to distinguish these two types of compact stars, quark stars and hadronic neutron stars. In order to assess the credibility of the models it would be interesting to assess them using features that can be observationally measured. The three observables M, R, and $\Lambda$ were selected as they represent the quantities directly accessible from current neutron star observations.

\subsection{Compactness–Tidal Deformability Relation}
To construct a physically consistent synthetic dataset, we apply an approach based on the empirical relation between stellar compactness and tidal deformability $\Lambda$. Such relations have been extensively investigated in the context of equation of state-independent relations among neutron stars ~\cite{YagiYunes2013,Maselli2013,YagiYunesReview2017,Godzieba2021}. Previous studies have shown that the compactness and tidal deformability satisfy an approximately universal relation that can be expressed as a polynomial in logarithmic values~\cite{{Maselli2013}} 
\begin{equation}
C=a_0+a_1\ln\Lambda+a_2(\ln\Lambda)^2
\label{Maseli-1}
\end{equation}
In the present work, we fit the relation in a form in which the tidal deformability is expressed as a function of compactness
\begin{equation}
\ln\Lambda
=a+b\ln C+c(\ln C)^2
\label{Maseli-2}
\end{equation}
The choice of this parametrization is motivated by the known relation between tidal deformability, compactness, and the Love number $k_2$ that is $\Lambda = \frac{2}{3} k_2 C^{-5}$.
This demonstrates that the relation between tidal deformability and compactness becomes approximately linear with deviations induced by the variation of the Love number $k_2$. Therefore, expressing $\ln\Lambda$ as a polynomial function of lnC provides a natural and physically motivated extension capable of capturing possible deviations from exact linearity arising from differences between stellar families and equations of state.
This representation provides a flexible approximation to the $\Lambda(C)$ relation and allows a implemented using linear regression on polynomial features. To improve the accuracy due to differences of the structural properties of hadronic and quark stars, we apply the fitting procedure independently for each stellar family.
The resulting coefficients and intrinsic scatter are reported in Table~\ref{tab:lambda_c_fit}.
\begin{table}[h]
\caption{Coefficients of the quadratic polynomial fit $\ln\Lambda = a + b\ln C + c(\ln C)^2$ for hadronic and quark stars. The table also shows the coefficient of determination $R^2$ and the intrinsic scatter $\sigma_\Lambda$ of the relation in log-space.}
\centering
\begin{tabular}{lccccc}
\hline
Category & $a$ & $b$ & $c$ & $R^2$ & $\sigma_\Lambda$ \\
\hline
Hadronic & -7.5629 & -8.7259 & -0.7987 & 0.9944 & 0.2595 \\
\hline
Quark    & -7.9265 & -9.7293 & -0.7960 & 0.9986 & 0.1319 \\
\hline
\end{tabular}
\label{tab:lambda_c_fit}
\end{table}

\subsection{Generating the Synthetic Data}
Following this procedure, we introduced three categories of uncertainties, namely mild, medium, and maximum, in the mass and radius measurements, corresponding to increasing noise levels, to evaluate whether the models remain robust under those conditions. 
More specifically for each $M,R$ pair of our initial dataset, we created four perturbed pairs according to Gaussian distributions, to create three different datasets corresponding to increasing errors mentioned above, with four times more perturbed data (150112) from the initial theoretical (37528 cases), for each uncertainty level. The reason is to achieve sufficient statistical coverage of the noise distribution while keeping computational cost manageable.  The three noise levels 
are summarized in Table~\ref{tab:noise_levels}. 
The perturbed masses $M'$ and radii $R'$ are given by:
\begin{equation}
M'= M\left(1 + \mathcal{N}(0,\sigma_M)\right), 
R'= R\left(1 + \mathcal{N}(0,\sigma_R)\right)
\end{equation}
\begin{table}[h]
\caption{Relative Gaussian uncertainties introduced in the stellar mass and radius when generating the synthetic datasets.}
\centering
\begin{tabular}{lcc}
\hline
Dataset & Mass uncertainty ($\sigma_M$) & Radius uncertainty ($\sigma_R$) \\
\hline
mild   & 2\%   & 4\%  \\
medium & 3.5\% & 7\%  \\
max    & 5\%   & 10\% \\
\hline
\end{tabular}
\label{tab:noise_levels}
\end{table}
The compactness of the perturbed configuration is then computed as $C'=GM'/R'c^2$ and the tidal deformability from the fitted compactness-deformability relation 
\begin{equation}
\ln\Lambda'=f(\ln C')+\epsilon
\label{Masseli-3}
\end{equation}
Using the original dataset, the fit for the compactness relation is performed through linear regression using polynomial features to calculate the $a$,$b$,$c$ terms and $\epsilon \sim N(0,\sigma_\Lambda)$ where $\sigma_\Lambda$ is the intrinsic scatter of the $\Lambda$-C relation in log-space. This procedure ensures that the generated datasets preserve the physical correlation between compactness and tidal deformability while including both uncertainties and internal variability of the EoS.

\section{Machine and Deep Learning Models Used for the Uncertainty-Injected Dataset}

In this section, we apply our models to the above three synthetic datasets based on simulated $M,R,\Lambda$ data. We use our initial ML models  Random Forest, Decision Tree, XGBoost, and Logistic Regression to evaluate the performance. For the neural network, we initially adopt the same strategy, and additionally, we create a new classification dataset using $M,R,\log\Lambda$ to improve our results by bringing all features in the same order of magnitude.

\subsection{Impact of synthetic data on Machine Learning Model Performance}

With the introduction of artificial errors in the physical quantities $(M, R, \Lambda)$, a gradual decrease in AUC and classification accuracy was observed across all models as the relative error increases from mild $\rightarrow$ medium $\rightarrow$ maximum (see also Table~\ref{tab:exhaustive_search_results} and Appendix for the unperturbed data AUC).
We first consider the mild-uncertainty dataset.
\begin{table}[h]
\caption{Performance metrics of the ML classifiers evaluated with mild errors.}
\centering
\begin{tabular}{lcccc}
\hline
\textbf{Metrics} & \textbf{RF} & \textbf{XGB} & \textbf{DT} & \textbf{LR} \\
\hline
Accuracy  & 0.995905 & 0.995938 & 0.989280 & 0.755027 \\
Precision & 0.995675 & 0.994208 & 0.989916 & 0.735584 \\
Recall    & 0.996382 & 0.997933 & 0.989276 & 0.819121 \\
F1-score  & 0.996029 & 0.996067 & 0.989596 & 0.775109 \\
AUC-score & 0.999823 & 0.999908 & 0.989280 & 0.825261 \\
\hline
\end{tabular}
\label{tab:mild_errors}
\end{table}

Among the examined models, XGBoost achieves the best overall performance, maintaining near-perfect results across most evaluation metrics, with an AUC value very close to unity. Random Forest also demonstrates highly robust classification performance, with accuracy, precision, recall, F1-score, and AUC values comparable to those of XGBoost.
Decision Tree presents slightly lower, but still very strong, performance, indicating that it remains effective under mild errors. In contrast, Logistic Regression shows the largest reduction in performance, with lower accuracy, precision, F1-score, and AUC compared to the tree-based models. This behavior suggests that Logistic Regression is more sensitive to stochastic perturbations in the input features.

\begin{table}[H]
\caption{Performance metrics of the ML classifiers evaluated  with medium errors.}
\centering
\begin{tabular}{lcccc}
\hline
\textbf{Metrics} & \textbf{RF} & \textbf{XGB} & \textbf{DT} & \textbf{LR} \\
\hline
Accuracy  & 0.994373 & 0.994873 & 0.987149 & 0.750067 \\
Precision & 0.993744 & 0.992797 & 0.986715 & 0.730740 \\
Recall    & 0.995349 & 0.997287 & 0.988372 & 0.815568 \\
F1-score  & 0.994546 & 0.995037 & 0.987543 & 0.770828 \\
AUC-score & 0.999771 & 0.999868 & 0.987110 & 0.819235 \\
\hline
\end{tabular}
\label{tab:medium_errors}
\end{table}
As shown in Table~\ref{tab:medium_errors}, the introduction of medium  noise leads to a slight reduction in the classification performance of the tree-based models compared to the mild error case. Nevertheless, both XGBoost and Random Forest maintain excellent predictive performance, with accuracy and AUC values remaining very close to unity.
The Decision Tree model shows a slight reduction compared to Random Forest and XGBoost, although its performance remains very strong across all evaluation metrics.
In contrast, Logistic Regression remains the most affected model, presenting substantially lower accuracy, precision, F1-score, and AUC values compared to the tree-based classifiers. 
\begin{table}[H]
\caption{Performance metrics of the ML classifiers evaluated with maximum errors.}
\centering
\begin{tabular}{lcccc}
\hline
\textbf{Metrics} & \textbf{RF} & \textbf{XGB} & \textbf{DT} & \textbf{LR} \\
\hline
Accuracy  & 0.992675 & 0.994373 & 0.985384 & 0.742343 \\
Precision & 0.992576 & 0.992727 & 0.985100 & 0.724104 \\
Recall    & 0.993217 & 0.996382 & 0.986563 & 0.807881 \\
F1-score  & 0.992896 & 0.994551 & 0.985831 & 0.763702 \\
AUC-score & 0.999747 & 0.999838 & 0.985347 & 0.810381 \\
\hline
\end{tabular}
\label{tab:max_errors}
\end{table}

As shown in Table~\ref{tab:max_errors}, the increase in noise to the maximum level leads to a further, but still limited, reduction in the performance of the tree-based classifiers. XGBoost remains the best-performing model overall, preserving near-perfect classification performance with an accuracy of 0.994373 and an AUC-score of 0.999838.
Random Forest also demonstrates strong robustness under maximum uncertainties, with accuracy, F1-score, and AUC-score values remaining very high. The Decision Tree model exhibits a more noticeable decrease compared to Random Forest and XGBoost, although it still maintains strong classification performance across all evaluation metrics.
In contrast, Logistic Regression remains the most sensitive model to noise, presenting the lowest accuracy, precision, F1-score, and AUC-score among all classifiers.

\begin{figure}[t]
\centering
\begin{subfigure}{0.43\textwidth}
\centering
\includegraphics[width=\linewidth]{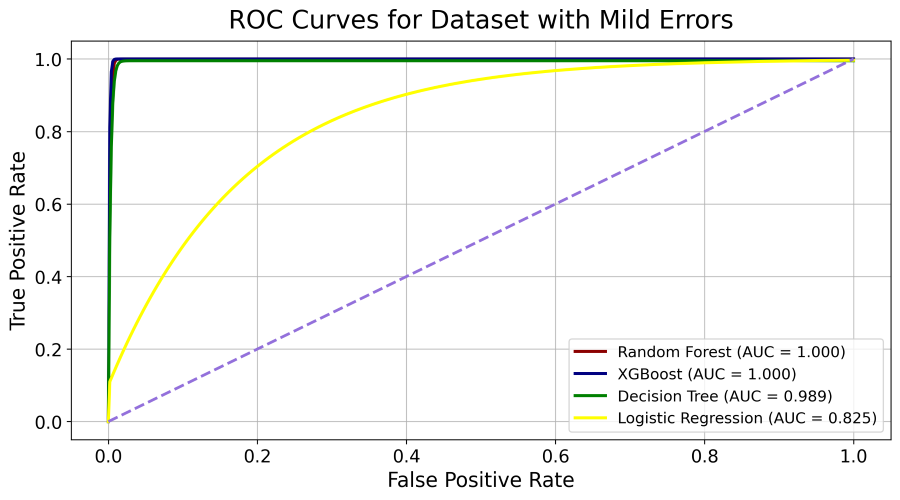}
\end{subfigure}
\caption{Receiver Operating Characteristic curves for the four ML classifiers. 
The upper panel corresponds to the original dataset based on theoretical calculations, while the lower panel shows the results obtained after introducing mild errors.}
\label{fig:roc_comparison_mild}
\end{figure}

Figure \ref{fig:roc_comparison_mild} presents the ROC curves for the four ML classifiers when mild errors are introduced into the dataset. In this scenario, small stochastic perturbations are applied to the stellar parameters, corresponding to $\sigma_M$ and $\sigma_R$ values representative of mild  uncertainties.
The Random Forest and XGBoost classifiers both achieve AUC values very close to 1.000, indicating an almost perfect discriminative capability. This result shows that the selected features still contain sufficient information to clearly distinguish between the two stellar classes, despite the presence of small errors.
The Decision Tree classifier also performs very well, with an AUC of 0.989. The slight decrease compared to the ensemble-based methods suggests a minor sensitivity to the stochastic dispersion of the data, although its classification ability remains highly robust.
The Logistic Regression model presents the lowest performance among the four classifiers, with an AUC of 0.825. This reduction indicates that the model is more affected by perturbations in the input features. 

\begin{figure}[t]
\centering
\begin{subfigure}{0.43\textwidth}
\centering
\includegraphics[width=\linewidth]{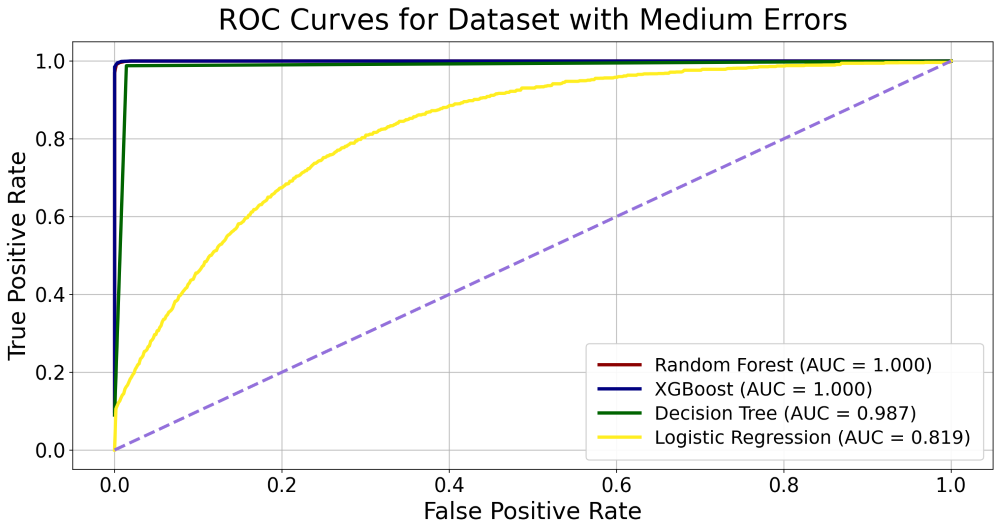}
\end{subfigure}
\caption{As in Figure~9, but for the dataset with medium errors.}
\label{fig:roc_comparison_medium}
\end{figure}

Figure \ref{fig:roc_comparison_medium} presents the ROC curves for the four ML classifiers when medium errors are introduced into the dataset. The results show that the discriminative capability of most models remains high, although a slight degradation is observed for some classifiers under increased stochastic perturbations.
Both Random Forest and XGBoost maintain an almost perfect classification performance, with AUC values very close to 1.000. This indicates that these models remain highly robust even in the presence of moderate uncertainties. The Decision Tree classifier also demonstrates very strong performance, achieving an AUC of 0.987.
In contrast, Logistic Regression exhibits the lowest performance among the four classifiers, with an AUC of 0.819. This result confirms that the model is more sensitive and is less effective in preserving a clear linear decision boundary between the two classes.

\begin{figure}[t]
\centering
\begin{subfigure}{0.43\textwidth}
\centering
\includegraphics[width=\linewidth]{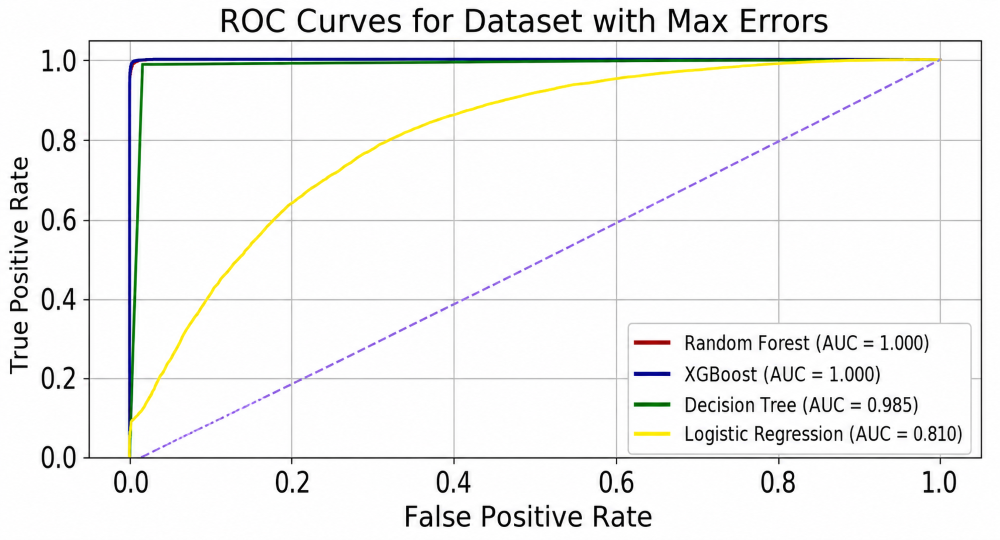}
\end{subfigure}

\caption{As in Figure~9, but for the dataset with maximum  errors.}
\label{fig:roc_comparison_max}
\end{figure}

Figure \ref{fig:roc_comparison_max} presents the ROC curves for the four ML classifiers when maximum  errors are introduced into the dataset. Under this highest level of stochastic perturbations, the overall discriminative capability of the models remains high, although a more noticeable reduction is observed for some classifiers.
Both Random Forest and XGBoost maintain an almost perfect classification performance, with AUC values very close to 1.000. The Decision Tree classifier also retains a very strong performance, with an AUC of 0.985. In contrast, Logistic Regression exhibits the lowest performance among the four classifiers, with an AUC of 0.810.

\begin{figure}[t]
\centering
\includegraphics[width=0.48\textwidth]{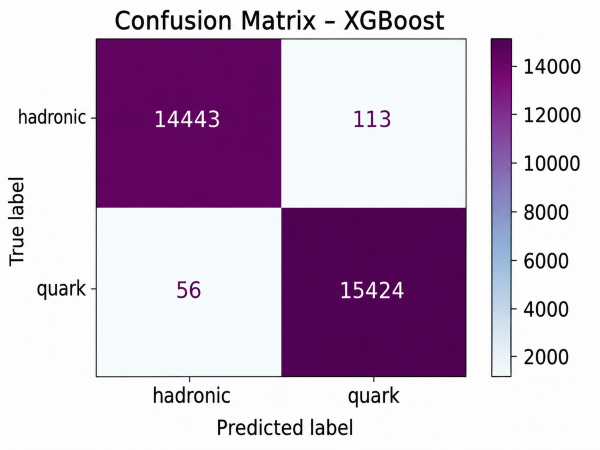}
\caption{Confusion matrix of the XGBoost classifier for the synthetic dataset with maximum errors using $(M, R, \Lambda)$. The small number of off-diagonal entries shows that the model remains highly accurate but not perfectly error-free under the strongest perturbations.}
\label{fig:confusion_matrix_XGB_max}
\end{figure}

To complement the ROC analysis, Figure~\ref{fig:confusion_matrix_XGB_max} presents the confusion matrix of the XGBoost classifier for the synthetic dataset with maximum errors. This case corresponds to the most demanding uncertainty scenario considered in this work. The classifier correctly identifies the vast majority of both hadronic and quark-star configurations, while only a small number of samples appear in the off-diagonal entries. More specifically, 113 hadronic configurations are classified as quark stars, whereas 56 quark-star configurations are classified as hadronic stars. This behavior confirms the robustness of the XGBoost model under strong perturbations, while also showing that the classification is not artificially perfect in the uncertainty-injected dataset.

\subsection{Impact of synthetic data on Neural Network Performance}

As discussed previously, three different uncertainty levels were considered, corresponding to mild, medium, and maximum errors. These datasets were generated in such a way that the physical correlation between compactness and tidal deformability was preserved.

First, Optuna, a Python library~\cite{Akiba-2019} for hyperparameter optimization, was applied in order to find the best neural-network architecture for the new feature set $(M,R,\Lambda)$. The Optuna-optimized neural network was also applied on the synthetic datasets with uncertainties. In this procedure, we obtained the following Optuna outputs corresponding to the best neural-network models for the three uncertainty levels:

\begin{itemize}
    \item \textbf{Mild errors:} two hidden layers with 80 and 32 neurons, dropout = 0.087, learning rate = $3.42 \times 10^{-4}$, batch size = 64, AdamW optimizer, and 116 epochs.
    
    \item \textbf{Medium errors:} two hidden layers with 112 and 112 neurons, dropout = 0.042, learning rate = $2.06 \times 10^{-4}$, batch size = 64, AdamW optimizer, and 155 epochs.
    
    \item \textbf{Maximum errors:} two hidden layers with 80 and 128 neurons, dropout = 0.010, learning rate = $2.35 \times 10^{-4}$, batch size = 64, AdamW optimizer, and 154 epochs.
\end{itemize}

Table~\ref{tab:nn_errors_mrlambda_optuna} presents the results for the mild, medium, and maximum error datasets using the feature set $(M,R,\Lambda)$, obtained with the best neural-network hyperparameters selected by Optuna for each corresponding uncertainty level. As expected, the classification performance decreases as the level of uncertainty increases. 

\begin{table}[h!]
\centering
\caption{Neural-network performance using Optuna for the synthetic datasets with uncertainties in the $(M,R,\Lambda)$ feature set.}
\begin{tabular}{lccc}
\hline
Metric & Mild Errors & Medium Errors & Max Errors \\
\hline
Accuracy  & 0.906279 & 0.895492 & 0.888467 \\
Precision & 0.871735 & 0.855464 & 0.865626 \\
Recall    & 0.959302 & 0.959302 & 0.927584 \\
F1-score  & 0.913425 & 0.904412 & 0.895534 \\
AUC-score & 0.976540 & 0.973821 & 0.969333 \\
\hline
\end{tabular}
\label{tab:nn_errors_mrlambda_optuna}
\end{table}

The relatively reduced performance of the Optuna-optimized model in the $(M, R, \Lambda)$ feature space can be attributed to the different numerical scales of the input variables. Neural networks are particularly sensitive to the relative scaling of the input features, since the training process depends on distances and gradients in the multidimensional feature space. In the present case, the stellar mass $M$ and radius $R$ vary within relatively limited numerical ranges, whereas the tidal deformability $\Lambda$ can span several orders of magnitude. This large scale separation may distort the effective geometry of the input space and make the optimization process less efficient.
To address this issue, the tidal deformability was transformed by using its logarithm, replacing $\Lambda$ with $\log\Lambda$. This transformation compresses the dynamical range of the tidal deformability and places it on a scale more comparable to the remaining input variables. Therefore, the neural network was trained and optimized again using the transformed feature set $(M,R,\log\Lambda)$.
The impact of this transformation is significant. The Optuna-optimized neural network using $(M,R,\log\Lambda)$ achieves almost perfect performance on the test set, with an accuracy of 0.9999, an F1-score of 0.9999, and an AUC-score equal to 1.0000.
The improvement remains equally clear when  uncertainties are included. Table~\ref{tab:nn_errors_mrloglambda_optuna} presents the results for the mild, medium, and maximum error datasets using the transformed feature set $(M,R,\log\Lambda)$. 

\begin{table}[h!]
\centering
\caption{Neural-network performance using Optuna for the synthetic datasets with uncertainties in the $(M,R,\log\Lambda)$ feature set.}
\begin{tabular}{lccc}
\hline
Metric & Mild Errors & Medium Errors & Max Errors \\
\hline
Accuracy  & 0.998435 & 0.998036 & 0.997503 \\
Precision & 0.998128 & 0.998062 & 0.997288 \\
Recall    & 0.998837 & 0.998127 & 0.997868 \\
F1-score  & 0.998482 & 0.998094 & 0.997578 \\
AUC-score & 0.999970 & 0.999969 & 0.999961 \\
\hline
\end{tabular}
\label{tab:nn_errors_mrloglambda_optuna}
\end{table}

These results demonstrate that the logarithmic transformation of the tidal deformability is essential for the efficient neural-network classification of compact stars when only the observationally motivated quantities are used. While the direct use of $\Lambda$ leads to a noticeable reduction in performance, the use of $\log\Lambda$ restores almost perfect classification accuracy and preserves the robustness of the model even under uncertainties. This confirms that the information contained in the tidal deformability is highly discriminative, provided that it is represented in a numerically suitable form for neural-network training.

\section{Results and Discussion}

Beyond the overall classification accuracy, the results provide important insights into the robustness of different ML algorithms when applied to astrophysical datasets. While all models achieved nearly perfect performance on the ideal, noise-free dataset, their behavior differed significantly when uncertainties were introduced.

Tree-based ensemble methods and particularly XGBoost, demonstrated remarkable robustness to stochastic perturbations in the input features. Even under the maximum noise scenario, the model maintained an AUC value very close to unity. This behavior can be attributed to the ensemble learning mechanism of gradient boosting, which combines multiple weak learners and effectively captures non-linear relationships in the feature space. As a result, the model remains capable of identifying subtle structural differences between neutron stars and quark stars even when the  measurements are partially degraded by noise.

In contrast, the Logistic Regression exhibited the highest sensitivity to the introduction of stochastic noise. This behavior is consistent with the linear nature of Logistic Regression, which relies on a simple linear decision boundary in the feature space. When  uncertainties increase, the overlap between the two stellar populations becomes larger, making it more difficult for a linear classifier to maintain a stable separation between the two classes. 

The Random Forest and Decision Tree classifiers display intermediate behavior. Although these models initially achieve almost perfect classification accuracy, their performance gradually deteriorates as the level of noise increases. This trend suggests that individual decision trees are more sensitive to fluctuations in the feature values compared to ensemble boosting approaches.

The exhaustive feature-subset search further confirms that not all physical quantities contribute equally to the classification task. Subsets containing the Love number $k_2$ consistently appear among the highest-ranked combinations, indicating that tidal-response information is particularly important for separating neutron stars from quark stars. The reduced subset $\{M,k_2,P_c\}$ provides an especially compact and robust representation, since it achieves excellent performance across the classical ML models and also preserves perfect predictive capability when used as input to the neural network.

An important physical implication of these results is that the selected observables contain strong and consistent information about the internal composition of compact stars. In particular, the tidal deformability and the Love number ($k_2$) appear to be highly informative quantities. These parameters are directly related to the response of a compact star to external tidal fields and therefore encode information about the stiffness of the underlying EoS.

Another important observation concerns the behavior of the neural network model under uncertainties. Although the neural network achieves perfect classification performance on the initial dataset, its accuracy decreases significantly when noise is introduced into the input features. However, the analysis of the training and validation loss curves shows that the degradation in performance is not caused by overfitting or instability during training. Instead, the loss curves converge smoothly and remain closely aligned throughout the training process.
This result suggests that the reduced accuracy originates from the intrinsic increase in the complexity of the classification problem. As stochastic noise is introduced into the physical parameters, the separation between neutron stars and quark stars in the multidimensional feature space becomes less distinct. 

Another important result concerns the treatment of the tidal deformability in the observationally motivated feature space. When the neural network is trained directly on $(M,R,\Lambda)$, the performance is noticeably reduced, mainly due to the different order of magnitude of $\Lambda$ compared to M,R. However, replacing $\Lambda$ with $\log\Lambda$ significantly improves the training process and restores almost perfect classification performance, even under uncertainties. This shows that the tidal deformability remains highly informative, provided that it is represented in a numerically suitable form for neural-network optimization.

The tree-based models used in this work remain highly robust under uncertainties. This behavior arises because Decision Tree rely on threshold-based splits rather than on the absolute scale of the input features. Consequently, moderate perturbations in the values of $M$, $R$, $\Lambda$ have only a limited impact on the learned decision boundaries. In contrast, neural networks are generally more sensitive to differences in feature magnitudes, making appropriate feature scaling, such as the use of $\log \Lambda$, beneficial for improving training stability and robustness.

From a broader astrophysical perspective, these findings highlight the importance of observational precision in constraining the equation of state of dense nuclear matter. Even relatively small uncertainties in key observables such as mass, radius, and tidal deformability can significantly affect the ability of ML models to distinguish between different classes of compact objects.

\section{Concluding Remarks}

The present study is motivated by the aim of employing statistical methods to enable the classification of astrophysical compact objects, whose fundamental properties are inferred from astrophysical observationswhich primarily concern neutron stars and the hypothetical class of quark stars. Such a classification is of considerable importance, as it provides the means not only to distinguish between these two types of objects on the basis of perturbed data, but also to assess the validity of existing theoretical models. 
In particular, the objective of this study is to investigate which physical quantities are most decisive for the identification and discrimination between the two aforementioned categories. Ultimately, this approach contributes to a deeper understanding of the composition and properties of dense nuclear matter.

From a technical perspective, the primary aim was to use these five microscopic and macroscopic features in order to accurately classify quark and hadronic neutron stars. At the same time, we sought to determine which of the five variables, mass, radius, tidal deformability, Love number, and central pressure, played the most significant role in the classification. To achieve this, we employed ML models, Decision Tree, Random Forest, and XGBoost, while  we also included Logistic Regression as an additional model. The exhaustive feature-subset search revealed a degree of degeneracy among the five input features, indicating that comparable classification performance could be achieved using only three of them. The three optimal features identified were $M,k_2,P_c$ but also other triplets such as {$M,\Lambda,k_2$} gave a very good accuracy (see the Appendix). To verify this, we applied the same process using a neural network and performed selective ablation of the original five features. We observed the same pattern, reinforcing that these three key variables were sufficient. 

As a next step, we sought a model relating directly to observable quantities, such as mass, radius, and tidal deformability. We created three error categories: mild, medium, and maximum. Applying the same approach to the three features, we observed a very minor decline in performance with increasing errors for the tree-based models. Decision Tree, Random Forest and XGBoost, while Logistic Regression dropped more significantly. This difference can be partly attributed to the threshold-based nature of tree models, which makes them less sensitive to the absolute scale of the input features. A similar behavior was observed for the neural network, which can be mainly attributed to the different orders of magnitude of the input features. To reduce the scale disparity among the input features, we trained the neural network using the feature set $(M, R, \log \Lambda)$, achieving near-perfect accuracy and confirming the beneficial effect of reducing differences in feature magnitudes. The main conclusion is that a degeneracy exists, and three-feature combinations can achieve near-optimal performance. The models handle errors well, which is crucial for classifying categories of compact objects from future observational data.

\appendix

\section{Additional figures and tables }
The Appendix provides the complete feature-subset rankings obtained from the exhaustive search procedure for each of the four ML classifiers. For completeness, Tables XII–XV present the full ranking of all 31 non-empty feature subsets, evaluated in terms of the mean ROC-AUC score and the corresponding standard deviation across the cross-validation folds. Table XII shows the results for the Random Forest classifier. Several feature combinations achieve essentially perfect performance, with the highest-ranked subset being \(\{M,P_c,k_2\}\). Similar behavior is observed for XGBoost in Table XIII, where the same subset also provides the best performance. Tables XIV and XV present the corresponding rankings for the Decision Tree and Logistic Regression classifiers, respectively. Although the exact ordering of the optimal subsets differs among the models, a common pattern emerges: feature combinations that include the Love number \(k_2\) consistently appear among the highest-ranked cases.


\begin{table*}
\caption{Random Forest feature subset ranking.}
\begin{tabular}{lllll}
\hline
Classifier & Feature Subset & Rank & Mean AUC & Std \\
\hline
RF & $\{M,P_c,k_2\}$ & 1 & 1.00000000 & 0.00000000 \\
RF & $\{M,k_2,\Lambda\}$ & 2 & 1.00000000 & 0.00000000 \\
RF & $\{M,R,P_c,k_2\}$ & 3 & 1.00000000 & 0.00000000 \\
RF & $\{M,P_c,k_2,\Lambda\}$ & 4 & 1.00000000 & 0.00000000 \\
RF & $\{R,P_c,k_2,\Lambda\}$ & 5 & 1.00000000 & 0.00000000 \\
RF & $\{M,R,P_c,k_2,\Lambda\}$ & 6 & 1.00000000 & 0.00000000 \\
RF & $\{k_2,\Lambda\}$ & 7 & 1.00000000 & 0.00000000 \\
RF & $\{R,k_2,\Lambda\}$ & 8 & 1.00000000 & 0.00000000 \\
RF & $\{P_c,k_2,\Lambda\}$ & 9 & 1.00000000 & 0.00000000 \\
RF & $\{M,R,k_2\}$ & 10 & 0.99999998 & 0.00000003 \\
RF & $\{R,P_c,k_2\}$ & 11 & 0.99999998 & 0.00000003 \\
RF & $\{M,R,k_2,\Lambda\}$ & 12 & 0.99999998 & 0.00000003 \\
RF & $\{M,R,P_c,\Lambda\}$ & 13 & 0.99999010 & 0.00001191 \\
RF & $\{M,R,\Lambda\}$ & 14 & 0.99998762 & 0.00000880 \\
RF & $\{M,R,P_c\}$ & 15 & 0.99991591 & 0.00002733 \\
RF & $\{P_c,k_2\}$ & 16 & 0.99898226 & 0.00017758 \\
RF & $\{M,k_2\}$ & 17 & 0.99858695 & 0.00029883 \\
RF & $\{R,P_c,\Lambda\}$ & 18 & 0.99597477 & 0.00024828 \\
RF & $\{R,k_2\}$ & 19 & 0.99063614 & 0.00052805 \\
RF & $\{M,P_c,\Lambda\}$ & 20 & 0.97809710 & 0.00237329 \\
RF & $\{k_2\}$ & 21 & 0.96788773 & 0.00164115 \\
RF & $\{M,R\}$ & 22 & 0.92260314 & 0.00125789 \\
RF & $\{R,P_c\}$ & 23 & 0.91128813 & 0.00290798 \\
RF & $\{R,\Lambda\}$ & 24 & 0.90965259 & 0.00371948 \\
RF & $\{M,\Lambda\}$ & 25 & 0.90696589 & 0.00124525 \\
RF & $\{M,P_c\}$ & 26 & 0.90628390 & 0.00194780 \\
RF & $\{P_c,\Lambda\}$ & 27 & 0.89256855 & 0.00211871 \\
RF & $\{R\}$ & 28 & 0.78739361 & 0.00252819 \\
RF & $\{\Lambda\}$ & 29 & 0.70261248 & 0.00443268 \\
RF & $\{P_c\}$ & 30 & 0.67145484 & 0.00573318 \\
RF & $\{M\}$ & 31 & 0.61053036 & 0.00593828 \\
\hline
\end{tabular}
\end{table*}

\begin{table*}
\caption{XGBoost feature subset ranking.}
\begin{tabular}{lllll}
\hline
\hline
Classifier & Feature Subset & Rank & Mean AUC & Std \\
\hline
XGB & $\{M,P_c,k_2\}$ & 1 & 1.00000000 & 0.00000000 \\
XGB & $\{M,R,P_c,k_2\}$ & 2 & 1.00000000 & 0.00000000 \\
XGB & $\{M,P_c,k_2,\Lambda\}$ & 3 & 1.00000000 & 0.00000000 \\
XGB & $\{M,R,P_c,k_2,\Lambda\}$ & 4 & 1.00000000 & 0.00000000 \\
XGB & $\{R,P_c,k_2,\Lambda\}$ & 5 & 1.00000000 & 0.00000000 \\
XGB & $\{M,R,k_2,\Lambda\}$ & 6 & 1.00000000 & 0.00000000 \\
XGB & $\{M,k_2,\Lambda\}$ & 7 & 0.99999998 & 0.00000003 \\
XGB & $\{P_c,k_2,\Lambda\}$ & 8 & 0.99999998 & 0.00000003 \\
XGB & $\{R,k_2,\Lambda\}$ & 9 & 0.99999995 & 0.00000009 \\
XGB & $\{R,P_c,k_2\}$ & 10 & 0.99999963 & 0.00000047 \\
XGB & $\{M,R,k_2\}$ & 11 & 0.99999957 & 0.00000018 \\
XGB & $\{M,R,P_c,\Lambda\}$ & 12 & 0.99980705 & 0.00007044 \\
XGB & $\{k_2,\Lambda\}$ & 13 & 0.99978579 & 0.00006077 \\
XGB & $\{M,k_2\}$ & 14 & 0.99780209 & 0.00029807 \\
XGB & $\{P_c,k_2\}$ & 15 & 0.99677489 & 0.00031049 \\
XGB & $\{M,R,\Lambda\}$ & 16 & 0.99647077 & 0.00083398 \\
XGB & $\{M,R,P_c\}$ & 17 & 0.99555741 & 0.00086737 \\
XGB & $\{R,k_2\}$ & 18 & 0.98733554 & 0.00061635 \\
XGB & $\{k_2\}$ & 19 & 0.98167664 & 0.00074621 \\
XGB & $\{R,P_c,\Lambda\}$ & 20 & 0.96439114 & 0.00208510 \\
XGB & $\{M,P_c,\Lambda\}$ & 21 & 0.91820773 & 0.00253326 \\
XGB & $\{R,\Lambda\}$ & 22 & 0.88873647 & 0.00302866 \\
XGB & $\{M,R\}$ & 23 & 0.87762456 & 0.00348757 \\
XGB & $\{R,P_c\}$ & 24 & 0.85490186 & 0.00332087 \\
XGB & $\{R\}$ & 25 & 0.83118888 & 0.00500576 \\
XGB & $\{P_c,\Lambda\}$ & 26 & 0.82797570 & 0.00290867 \\
XGB & $\{M,\Lambda\}$ & 27 & 0.82616669 & 0.00449058 \\
XGB & $\{\Lambda\}$ & 28 & 0.78228371 & 0.00490336 \\
XGB & $\{M,P_c\}$ & 29 & 0.72233529 & 0.00313013 \\
XGB & $\{M\}$ & 30 & 0.63228843 & 0.00565419 \\
XGB & $\{P_c\}$ & 31 & 0.63061198 & 0.00299304 \\
\hline
\end{tabular}
\end{table*}

\begin{table*}
\caption{Decision Tree feature subset ranking.}
\begin{tabular}{lllll}
\hline
Classifier & Feature Subset & Rank & Mean AUC & Std \\
\hline
DT & $\{P_c,k_2,\Lambda\}$ & 1 & 0.99986654 & 0.00013364 \\
DT & $\{k_2,\Lambda\}$ & 2 & 0.99983440 & 0.00017369 \\
DT & $\{M,k_2,\Lambda\}$ & 3 & 0.99980235 & 0.00015045 \\
DT & $\{M,P_c,k_2,\Lambda\}$ & 4 & 0.99980235 & 0.00015045 \\
DT & $\{M,R,P_c,k_2,\Lambda\}$ & 5 & 0.99980235 & 0.00015045 \\
DT & $\{R,P_c,k_2,\Lambda\}$ & 6 & 0.99979988 & 0.00015149 \\
DT & $\{M,R,k_2,\Lambda\}$ & 7 & 0.99977022 & 0.00019769 \\
DT & $\{R,k_2,\Lambda\}$ & 8 & 0.99973825 & 0.00020940 \\
DT & $\{M,R,P_c,k_2\}$ & 9 & 0.99930830 & 0.00021964 \\
DT & $\{M,P_c,k_2\}$ & 10 & 0.99924164 & 0.00027337 \\
DT & $\{R,P_c,k_2\}$ & 11 & 0.99890303 & 0.00029056 \\
DT & $\{M,R,k_2\}$ & 12 & 0.99881771 & 0.00035115 \\
DT & $\{M,k_2\}$ & 13 & 0.99671814 & 0.00019278 \\
DT & $\{P_c,k_2\}$ & 14 & 0.99424111 & 0.00103359 \\
DT & $\{R,k_2\}$ & 15 & 0.98897541 & 0.00038336 \\
DT & $\{M,R,P_c\}$ & 16 & 0.98786988 & 0.00155186 \\
DT & $\{M,R,P_c,\Lambda\}$ & 17 & 0.98784406 & 0.00146851 \\
DT & $\{M,R,\Lambda\}$ & 18 & 0.98323151 & 0.00075005 \\
DT & $\{k_2\}$ & 19 & 0.97685173 & 0.00043647 \\
DT & $\{R,P_c,\Lambda\}$ & 20 & 0.94708872 & 0.00369272 \\
DT & $\{R,\Lambda\}$ & 21 & 0.89734947 & 0.00439956 \\
DT & $\{R,P_c\}$ & 22 & 0.89398673 & 0.00368875 \\
DT & $\{M,R\}$ & 23 & 0.89216202 & 0.00275002 \\
DT & $\{M,P_c,\Lambda\}$ & 24 & 0.82774916 & 0.00546569 \\
DT & $\{R\}$ & 25 & 0.81764481 & 0.00488823 \\
DT & $\{P_c,\Lambda\}$ & 26 & 0.81141444 & 0.00773070 \\
DT & $\{M,\Lambda\}$ & 27 & 0.76494400 & 0.02722395 \\
DT & $\{M,P_c\}$ & 28 & 0.75353785 & 0.00536002 \\
DT & $\{\Lambda\}$ & 29 & 0.70150535 & 0.01647151 \\
DT & $\{P_c\}$ & 30 & 0.63882017 & 0.00258723 \\
DT & $\{M\}$ & 31 & 0.61200354 & 0.00767090 \\
\hline
\end{tabular}
\end{table*}

\begin{table*}
\caption{Logistic Regression feature subset ranking.}
\begin{tabular}{lllll}
\hline
Classifier & Feature Subset & Rank & Mean AUC & Std \\
\hline
LR & $\{M,R,k_2\}$ & 1 & 1.00000000 & 0.00000000 \\
LR & $\{M,R,P_c,k_2\}$ & 2 & 1.00000000 & 0.00000000 \\
LR & $\{M,R,k_2,\Lambda\}$ & 3 & 1.00000000 & 0.00000000 \\
LR & $\{M,R,P_c,k_2,\Lambda\}$ & 4 & 0.99999982 & 0.00000031 \\
LR & $\{M,P_c,k_2,\Lambda\}$ & 5 & 0.99654229 & 0.00022567 \\
LR & $\{M,P_c,k_2\}$ & 6 & 0.99653950 & 0.00022698 \\
LR & $\{M,k_2,\Lambda\}$ & 7 & 0.98264876 & 0.00090938 \\
LR & $\{M,k_2\}$ & 8 & 0.98262657 & 0.00091089 \\
LR & $\{P_c,k_2,\Lambda\}$ & 9 & 0.98214949 & 0.00049132 \\
LR & $\{P_c,k_2\}$ & 10 & 0.98204225 & 0.00050090 \\
LR & $\{R,P_c,k_2,\Lambda\}$ & 11 & 0.98079513 & 0.00068436 \\
LR & $\{R,P_c,k_2\}$ & 12 & 0.98064353 & 0.00070005 \\
LR & $\{R,k_2,\Lambda\}$ & 13 & 0.96408569 & 0.00123215 \\
LR & $\{R,k_2\}$ & 14 & 0.96366927 & 0.00126806 \\
LR & $\{k_2,\Lambda\}$ & 15 & 0.95208217 & 0.00099657 \\
LR & $\{k_2\}$ & 16 & 0.95130746 & 0.00104584 \\
LR & $\{M,R,P_c,\Lambda\}$ & 17 & 0.93286802 & 0.00193464 \\
LR & $\{M,R,P_c\}$ & 18 & 0.93226671 & 0.00195030 \\
LR & $\{M,R,\Lambda\}$ & 19 & 0.84836675 & 0.00179339 \\
LR & $\{M,R\}$ & 20 & 0.84662263 & 0.00181486 \\
LR & $\{R,P_c\}$ & 21 & 0.82437436 & 0.00445738 \\
LR & $\{R,P_c,\Lambda\}$ & 22 & 0.82396213 & 0.00444787 \\
LR & $\{R\}$ & 23 & 0.81670803 & 0.00327994 \\
LR & $\{R,\Lambda\}$ & 24 & 0.81595003 & 0.00330409 \\
LR & $\{M,\Lambda\}$ & 25 & 0.51749328 & 0.00428820 \\
LR & $\{M,P_c,\Lambda\}$ & 26 & 0.51527404 & 0.00333814 \\
LR & $\{M\}$ & 27 & 0.51431272 & 0.00867817 \\
LR & $\{\Lambda\}$ & 28 & 0.50888664 & 0.00672950 \\
LR & $\{P_c,\Lambda\}$ & 29 & 0.48243243 & 0.00470882 \\
LR & $\{P_c\}$ & 30 & 0.44869974 & 0.00457618 \\
LR & $\{M,P_c\}$ & 31 & 0.43788422 & 0.00456395 \\
\hline
\end{tabular}
\end{table*}

\FloatBarrier

\end{document}